# Size dependence of nanoscale wear of silicon carbide

Chaiyapat Tangpatjaroen[1], David Grierson[1], Steve Shannon[2], Joseph E. Jakes[3], Izabela Szlufarska[1,*]

[1]Department of Materials Science & Engineering, University of Wisconsin—Madison, Madison, Wisconsin 53706, United States

[2]Department of Nuclear Engineering, North Carolina State University, Raleigh, North Carolina 27695, United States

[3]Forest Biopolymers Science and Engineering, USDA Forest Service, Forest Products Laboratory. One Gifford Pinchot Drive, Madison, Wisconsin 53726, United States

Corresponding author's email: szlufarska@wisc.edu

**ABSTRACT**

Nanoscale, single-asperity wear of single-crystal silicon carbide (sc-SiC) and nanocrystalline silicon carbide (nc-SiC) is investigated using single-crystal diamond nanoindenter tips and nanocrystalline diamond atomic force microscopy (AFM) tips under dry conditions, and the wear behavior is compared to that of single-crystal silicon with both thin and thick native oxide layers. We discover a transition in the relative wear resistance of the SiC samples compared to that of Si as a function of contact size. With larger nanoindenter tips (tip radius ~370 nm), the wear resistances of both sc-SiC and nc-SiC are higher than that of Si. This result is expected from the Archard's equation because SiC is harder than Si. However, with the smaller AFM tips (tip radius ~20 nm), the wear resistances of sc-SiC and nc-SiC are lower than that of Si, despite the fact that the contact pressures are comparable to those applied with the nanoindenter tips, and the plastic zones are well developed in both sets of wear experiments. We attribute the decrease in the relative wear resistance of SiC compared to that of Si to a transition from a wear regime dominated by the materials' resistance to plastic deformation (i.e., hardness) to a regime dominated by the materials' resistance to interfacial shear. This conclusion is supported by our AFM studies of wearless friction which reveal that the interfacial shear strength of SiC is higher than that of Si. The contributions of surface roughness and surface chemistry to differences in interfacial shear strength are also discussed.

**INTRODUCTION**

Undesirable friction and uncontrolled wear can be particularly critical in nanoscale contacts because of the high surface-to-volume ratio and these phenomena can severely limit durability and usability of micro/nanoscale devices. Examples include failure due to stiction (i.e., large adhesion) in micro/nano electromechanical systems (MEMS/NEMS) [1-2] and wear of nanometer-sized scanning tips in magnetic storage devices.[3] Moreover, tribological properties derived from traditional testing methods and models are often insufficient for characterizing material behavior at small scales, and, in such cases, experiments and simulations are needed at the scale of interest in order to determine the parameters and the mechanisms that are relevant to the design of such small-scale devices.[4-6]

Friction and wear studies of small-scale mechanical contacts can reveal new and interesting phenomena that are not observed at larger scales, uncovering new relationships that cannot be described by conventional models. For instance, in small contacts in the wearless regime, continuum contact mechanics and the corresponding friction laws have been shown to break down,[7-8] and contacts between certain materials have been shown to exhibit a negative coefficient of friction.[9] Nanoscale contacts can also display unique mechanisms of wear, such as the removal of one atom at a time from the surface as it wears.[10-11] This atom-by-atom removal mechanism can be operative in small contacts at relatively low normal loads. For ceramics and semiconductors, reducing the contact size can be particularly advantageous because nanometer-sized cutting tools can suppress sliding-induced fracture, and the nominally brittle ceramics can be machined in a ductile manner.[12-13] An interesting question that arises from these experiments is whether the plastic wear of ceramics, which is brought about by small contact sizes, follows the same qualitative trends as the wear of metals. In addition, it is unknown whether harder materials will exhibit a better wear resistance at the nanoscale; such behavior would be expected from the Archard's equation, which has been validated mainly for macro and micro-scale contacts.[14]

In this study, we focus on silicon carbide (SiC) because of its many outstanding properties, such as high hardness and stiffness, corrosion resistance, stability at high temperature, large bandgap, high breakdown electric field, and the fact that SiC can be manufactured as a thin film coating.[15-19] Among its various uses, SiC is considered one of the prime candidates to replace Si in the MEMS devices for harsh environments due to the superior mechanical and electrical properties of SiC at high temperatures.[20-22] The feasibility of fabricating SiC for MEMS has been demonstrated by Roy *et al.*,[23] whereas the first SiC-based NEMS was synthesized by Yang *et al.*[24-25] for the purpose of building a high-frequency nanometer-scale resonator. SiC has also been shown to be successful as a tribological coating material for MEMS. For example, a polycrystalline SiC coating was used by Gao *et al.*[26] to reduce adhesion in Si-based MEMS. Similarly, Ashurst *et al.*[27] have shown that under the contact pressure of 78 MPa, a SiC coating in the sidewall friction tester has less wear scarring than other MEMS materials.

In addition to single-crystal SiC (sc-SiC), we also consider a nanocrystalline SiC (nc-SiC) due to the fact that nc materials have been often found to have higher hardness and wear resistance than their coarse-grained counterparts.[28-29] In fact, both simulations[30-31] and experimental studies of nc-SiC deposited by thermal plasma chemical vapor deposition[32] found that the hardness and strength of this material can exceed the corresponding values of single crystal SiC.

Several tribological studies have been performed on SiC at the macroscale.[33-35] As expected from the high hardness of the material, SiC was found to exhibit excellent tribological properties, such as the wear resistance. Similar results have been found at the microscale. For instance, studies using a microtribometer found that SiC has a higher wear resistance and generates fewer wear debris particles as compared to CoCrMo, Ti-6Al-4V, and stainless steel.[36] In another example, Pöhlmann et al.[37] used a diamond tip to investigate the scratch behavior of single-crystal SiC at a normal load of 60 µN. It was found that Si, $SiO_2$, and oxidized SiC are less wear resistant than SiC.

Studies of the tribological properties of SiC at the nanoscale have been much scarcer. For instance, in the regime of low normal loads, where wear takes place through atom-by-atom removal,[10] nanometer-sized scanning probe microscopy tips made of amorphous SiC were shown to exhibit much less wear than tips made of Si. Zum Gahr et al.[38] investigated wear of SiC at the macro- and nanoscales. At the macroscale, the authors used a ball-on-disk setup to test SiC balls with radii of 1.6 mm and 10 mm, and they found that the wear and the friction coefficients depend on relative humidity (RH). At the nanoscale, the authors used Si AFM tips with a radius of curvature of 24 nm, and they found that the friction coefficient of SiC was nearly independent of RH. However, friction was only investigated at a load of 87.5 nN, and wear from the nanoscale test was not discussed in the paper. Sundararajan et al.[39] performed AFM studies with a diamond tip (radius of ~70 nm) using loads between 20 µN and 100 µN. They found that SiC has a higher wear resistance (i.e., a lower wear rate) than Si and $SiO_2$. In addition, the friction coefficient of a 3C-SiC film at low loads (50-300nN) was studied with a $Si_3N_4$ tip (radius of ~50 nm) under ambient conditions, and it was found that the friction coefficient of SiC is lower than that of Si. However, wear was not investigated in that load regime.

Up to this point, wear analysis on SiC has not been yet carried out within the load range that is relevant for MEMS/NEMS devices (in the order of 100 and 1000 nN). This is also the load range within which even a brittle material like SiC is expected to exhibit plastic deformation with a nanoscale contact point;[13] therefore, wear and friction measurements can be analyzed to address questions that are both fundamental and of practical importance, including whether or not wear resistance in nanoscale contacts is correlated with hardness. To address this question in particular, here we perform single-asperity wear tests on sc-SiC samples and nc-SiC thin films over the load range of 120 nN – 2 mN using tips with radii of curvature on the order of tens of nanometers and on the order of hundreds of nanometers. We compare the SiC wear results to those from two different single-crystal Si samples - one with a thin native oxide (labeled as Si), and the other with a thick (~16 nm) oxide (labeled as $SiO_x$).

**EXPERIMENTAL PROCEDURES**

**Materials and sample preparation.**

In this study we used single-crystal silicon carbide (sc-SiC) wafers, nanocrystalline silicon carbide (nc-SiC) thin films grown on silicon substrates, and single-crystal silicon (Si). The single-crystal 4H-SiC wafers were manufactured by CREE® with an orientation of (0001) ± 0.5, and the nanocrystalline 3C-SiC films with {111} texture were grown using CVD. The details of the process have been described elsewhere.[40] For this work, the as-deposited nc-SiC film had a thickness of 500 nm, and it exhibited a columnar grain structure with grains approximately 100 nm in width. Single-crystal silicon (Si) wafers (p-type, manufactured by Nova electronic materials, LLC) with (100) orientation were used as a reference material, with both thin and thick native oxides, in all experimental measurements. Prior to nanotribological measurements, the materials' surfaces were prepared as follows. For sc-SiC, the samples were mechanically polished using 30 µm, 15 µm, 6 µm, 1 µm, and 0.1 µm diamond papers, in that order. After mechanical polishing, the samples underwent three cycles of ultrasonic rinsing with acetone and methanol for 15 minutes per solvent. The final finishing step involved ion milling at 1 kV for 20 minutes. The final roughness achieved was around 0.2 nm ($R_q$), as measured over an area of 300×300 nm² using a Multimode 8 atomic force microscope (AFM) with a Nanoscope V controller (Bruker Corporation). Due to the thinness of the nc-SiC films, the surfaces of those samples were prepared solely by ion milling at 1 kV for 20 minutes. The lowest achieved roughness of the sample was 0.3 nm ($R_q$) over an area of 300×300 nm². The surfaces of the Si wafers were prepared by two methods. The first method was used to prepare a thick native oxide layer - the as-received wafer was only washed with acetone and methanol for 15 minutes. The sample with the thicker oxide was labeled $SiO_x$. In the second method, a chemical cleaning process was used to remove the thick native oxide layer. First, the as-received wafer was washed with acetone and methanol for 15 minutes. The wafer was then exposed to UV ozone for 20 minutes, rinsed with deionized water for 5 minutes, dipped in 10% hydrofluoric acid for 20 seconds, and then rinsed again with deionized water for another 5 minutes. The process starting from the UV ozone step was done twice to ensure smoothness and to remove contamination. The final finishing step was ion milling at 1 kV for 20 minutes. The level of roughness achieved by this process was 0.1 nm ($R_q$) over a 300×300 nm² area as measured by AFM.

**Scratch test.**

Two sets of scratch tests were performed on each sample. In the first set, we used a TI 950 TriboIndenter (Hysitron, Inc.) with a diamond Berkovich tip (measured tip radius of 370 nm). In the second set, we used the Multimode AFM with nanocrystalline diamond AFM tips (ND-DYIRS-5, K-Tek Nanotechnology, LLC, with tip radii measured between 11 nm and 27 nm). All the experiments were done in a dry nitrogen environment with a relative humidity less than 3±3%, as measured by a dew point thermo-hygrometer (Tech Instrumentation, Inc.). The radius of the Berkovich tip was calculated using a custom Matlab script and a least-squares algorithm to fit a series of parabolas to a top-down 3D AFM topography image of the nanoindenter tip taken with a sharp Si tip (NTESPA, Bruker Corporation) in tapping mode. Based on calibrations of the tip area function for the nanoindenter tip, the topmost 40 nm of the tip was chosen as the vertical range over which the parabolas were fit, and parabolas were fit in 5° increments to examine all possible profiles of the tip. The minimum value from the fitting routine was taken to be the radius of the tip. To ensure stability of the Berkovich tip during the experiment, the hardness of a reference sample was measured before and after scratch testing. The radii of the AFM tips were periodically characterized throughout the experiments using a Si TC1 calibration sample (Ted Pella, Inc) and the SPIP program (Image Metrology A/S) to reconstruct the tip shape. Calibration of the lateral forces in the AFM test was done using the vertical sidewall approach described in Ref.[41] For the scratch tests with the Berkovich tip, we formed 3 sets of 6 parallel scratches with different normal loads. Each scratch was 3 μm long and was made by carrying out 500 reciprocal scratch cycles. For the AFM experiments, we formed 3 sets of 7 scratches at different normal loads. Each scratch was 3 μm long and was made by carrying out 2400 reciprocal scratch cycles. After the scratch tests, a sharp tapping-mode Si AFM tip was used to characterize morphology of the scratches and the debris. The samples were then cleaned using First contact™ (Photonic Cleaning Technologies, LLC) followed by an ultrasonic bath in acetone and then in methanol to remove the wear debris. After debris removal, tapping-mode imaging was again performed on the scratches, and the maximum scratch depth and the groove volume were calculated from those images. The maximum scratch depth is measured relative to the average surface height. Cross-sectioning by focused ion beam (FIB) milling (Auriga, Carl Zeiss Microscopy) was also performed to determine the scratch depth for selected scratched more accurately when wear debris could not be removed via cleaning. Some of the maximum scratch depth values have been corrected by these FIB results, which have been explained earlier.

Each value of the maximum scratch depth, groove volume and the friction force measured with the Berkovich tip represents the average of measurements from three identical tests. The error bars indicate the calculated standard deviation between each test. Friction force in AFM is calculated as the average over 2,400 reciprocal scratch cycles performed at a given normal load. Error bar in friction force corresponds to the standard deviation from this average. Maximum scratch depth and groove volume in AFM are measured at the end of the 2,400 cycles and therefore there is no measurement of the spread (error bar) in this data.

**Wearless regime measurements.**

Friction vs load in the wearless regime was measured using a nanocrystalline diamond AFM tip (ND-CTIR2M-5, K-Tek Nanotechnology, LLC). The lateral force calibration procedure was done using a wedge-shaped Si calibration sample (TGG1, NT-MDT) with the method described in Ref.[42] AFM tip radii were characterized periodically using the TC1 sample mentioned previously along with the SPIP software. The applied load was varied up to 120 nN by varying the set point voltage using a function generator (33120A, Hewlett Packard Company). The tilt compensation technique[41, 43] was employed to keep the AFM scanning on a single line as the load was varied during the experiments. Tapping-mode imaging before and after each test was performed to ensure that the friction vs load experiments did not cause damage to the surface. Static pull-off force measurements were also performed before and after the friction tests. Each pull-off force reported represents the average of five measurements, and the error bars corresponded to the standard deviation from that average. Friction vs load curves were acquired from at least eight different locations on each sample. The interfacial shear strength was obtained from fitting the friction vs load data to the Maugis-Dugdale model[44] using the Carpick-Ogletree-Salmeron (COS) equation.[45] Input parameters in the model include the elastic modulus in Table 1, the radius of curvature of the tip in Table 2, and the Poisson's ratios which were previously shown. The plot of Figure 5a shows the curves that most accurately reflect the average interfacial shear strengths plotted in Figure 5b. The error bars represent the total standard deviation from all error values of interfacial shear strength measurement.

### Materials characterization

The surface chemistry and oxide thickness of each sample was characterized by X-ray photoelectron spectroscopy (K-Alpha, Thermo Fisher Scientific, Inc.) within one week after the samples were prepared. One week was also the wait time between sample preparation and scratch testing. For measuring oxide thickness we used the sputtering rate that had been determined from XPS experiments performed previously on Si sample with a thicker thermally grown oxide whose thickness was measured directly via SEM cross-sectional imaging. The hardnesses and elastic moduli of Si, sc-SiC, and the nc-SiC thin film were assessed using the Hysitron TI 950 TriboIndenter equipped with a Berkovich probe. A series of 150 nanoindents, with a 5 s load time, a 2 s hold at maximum load, and a 5 s unload time, with maximum loads ranging from 0.5 to 14 mN, were placed on the prepared surface of each material. A similar series of nanoindents were placed on a fused silica calibration standard to calibrate the machine compliance[46] and tip area function[47]. The hardnesses of all materials and the elastic moduli of Si and sc-SiC were measured using the Oliver-Pharr method.[47] Poisson's ratios of 0.28, 0.183, and 0.183 were used in the elastic modulus calculation for Si, sc-SiC, and nc-SiC, respectively. The nc-SiC thin-film modulus was determined by comparing experimental results to Stone's theoretical simulations[48] using previously described methods.[49-50] AFM imaging of residual nanoindent impressions revealed that while both the Si and sc-SiC nanoindents had the expected equilateral triangle shape, the nc-SiC nanoindents did not, indicating a tilted surface. The nc-SiC projected contact areas were corrected using a surface tilt correction based on the measured side lengths from the residual nanoindent impression.[51] As hardness and elastic moduli for sc-SiC and Si were found to be independent of the normal load, the reported mechanical properties correspond to the average over multiple measurements (128 points for sc-SiC and 134 points for Si). For nc-SiC, due to the fact that its hardness is depth dependent, we report an average over 10 points measured at the contact depth around 122 nm on 432 nm thickness nc-SiC sample (The thickness of the samples that were used for scratch experiments is ~310 nm). The error bars represent the standard deviation from the average values of hardness and elastic modulus.

**RESULTS AND DISCUSSION**

We first carried out wear tests using a Berkovich diamond nanoindenter with a radius of curvature of approximately 370 nm. We performed reciprocal scratch tests (500 scratch cycles at a rate of 2 Hz, each 3 μm in length) on sc-SiC, nc-SiC, Si, and $SiO_x$ in the load range of 50 – 2000 μN. The tests were carried out in a dry nitrogen atmosphere (RH < 3±3%). Figure 1a shows representative examples of AFM topography images of scratches on sc-SiC, nc-SiC, Si, and $SiO_x$ after 500 cycles of scratching under 1 mN applied load. These images were acquired using a sharp Si AFM tip, with a radius between 8-12 nm, after the scratch experiments were completed in order to obtain high-resolution of images of the worn areas for subsequent analysis. Tapping mode AFM was used for the imaging to minimize disturbance of wear debris.

In order to quantify wear resistance from the scratch tests, we primarily focused on two properties: the maximum scratch depth and the groove volume. The maximum scratch depth was measured by finding the minimum depth within the scratch relative to the average surface height outside of the scratch. We also cross-sectioned the wear track under selected scratch conditions using the focused ion beam (FIB) technique for cases in which the wear debris became trapped in the groove and could not be removed during surface cleaning. Specifically, we performed FIB for those conditions in which the trends in the maximum scratch depth and in the measured debris volume were inconsistent with each other. Specimens prepared by FIB were then analyzed in the scanning electron microscope to obtain more accurate values of the maximum scratch depth. In cases where both AFM imaging and SEM cross-sectional imaging were performed, the larger value for scratch depth was chosen to represent the maximum scratch depth.

Figure 1b shows the AFM profiles of the scratches taken through the middle of each scratch shown in Figure 1a. The profiles were shifted vertically for display purposes. Figures 1a and 1b reveal that sc-SiC and nc-SiC have lower scratch depths (and therefore better wear resistances) compared to those of Si and $SiO_x$ for the same scratching conditions. As shown in Figure 1c, the trend in scratch depth persists across the entire load range of 50 – 2000 μN. Looking at the thicknesses of the oxides present on the samples' surfaces (shown in Table 1), it is clear that the depths of the scratches in all four samples exceed the thicknesses of their oxide layers at the higher loads.

In this experiment, the applied contact pressures ranged from the elastic to the plastic deformation regimes. The mean contact pressure, $p_m$, was calculated using the combined effects of the average normal stress (p) and the tangential stress (τ) according to the following relation[52]

$$p_m = \min\left\{\sqrt{p^2 + \alpha\tau^2}, H\right\} \qquad (1)$$

where $H$ is the material's hardness and α = 9 is an empirical constant. The average normal stress, p, which was calculated using the measured applied force and Hertz equation.[52] The tangential stress, τ, was calculated using the measured friction force divided by the contact area. In Figure 1c, open symbols correspond to mean contact pressures that are lower than the hardness of the material, and solid symbols correspond to points where the mean contact pressure is equal to hardness, and plastic deformation is expected. The hardness values of sc-SiC, nc-SiC, and Si were obtained from load-displacement curves using standard nanoindentation techniques. Interestingly, we found that nc-SiC, in contrast to sc-SiC, has a depth dependent hardness (shown in supplementary Figure S1b). The depth dependence will not be analyzed in detail in this paper. Therefore, for nc-SiC, the hardness was taken to be the average hardness measured around the contact depth of 122 nm on ~432 nm nc-SiC at which the scratch experiments were performed on ~310 nm nc-SiC. The average hardness values of sc-SiC, nc-SiC, and Si were found to be 37±1 GPa, 26±2 GPa, and 12.6±0.2 GPa, respectively. The hardness of $SiO_x$ was taken to be equal to 8 GPa, which is the hardness of silicon oxide ($SiO_2$) reported in the literature.[47] Because our $SiO_x$ sample consists of a ~16 nm thick silicon oxide layer on silicon, the hardness of $SiO_x$ should be between 8 GPa (the value for $SiO_2$) and 12.6 GPa (the value for Si), depending on the depth of the scratch. Interestingly, the nc-SiC studied here has a lower hardness than that of sc-SiC, which is in contrast to findings from several simulations and experiments found in the literature.[30-32] A likely reason for this difference is the morphology of the grains. Unlike in the earlier studies, the nc-SiC studied here has columnar grains and a strong {111} texture. This result provides further evidence that when considering mechanical properties, one should consider the effects of grain orientation and grain boundary structure in addition to the effects of the grain size.[53-54]

**Table 1.** Mechanical and surface properties of the samples. RMS is an abbreviation for root mean squared.

| Sample | Hardness (GPa) | RMS roughness ($R_q$, nm) measured over area $A$ | | Oxide thickness (nm) | Type of oxide | Elastic modulus (GPa) |
| --- | --- | --- | --- | --- | --- | --- |
| | | $A = 0.09\ \mu m^2$ | $A = 20\ \mu m^2$ | | | |
| sc-SiC | 37±1 | 0.19±0.08 | 0.782±0.09 | 3.2 | $SiO_2$, $Si_4C_{4-x}O_2$ | 397±8.5 |
| nc-SiC | 26±2 | 0.58±0.26 | 3.36±0.45 | 1.8 | $SiO_2$, $Si_4C_{4-x}O_2$ | 292±36 |
| Si | 12.6±0.2 | 0.070±0.002 | 0.090±0.001 | 3.6 | $SiO_2$ | 160±5 |
| $SiO_x$ | $8^{47}$ | 0.16±0.01 | 0.28±0.18 | 16.2 | $SiO_2$ | $73^{47}$ |

The trend in wear resistance shown in Figure 1 can be explained based on the hardness of the materials under study. Wear resistance, which is inversely proportional to the wear rate, has been shown in many cases to be proportional to the hardness of a given material.[55-56] This empirical observation can be described by the Archard's equation,[57] which relates hardness to the worn volume during adhesive and abrasive wear of plastic contacts as follows:

$$V = \frac{kWx}{H} \qquad (2)$$

Here, $V$ is the total volume of debris formed during wear (or, equivalently, the volume of material removed), $k$ is a dimensionless wear coefficient, $W$ is the normal load, $x$ is the sliding distance, and $H$ is the hardness of the softer of the two contacting materials. As shown in Table 1, the hardness values of the materials decrease in order of sc-SiC, nc-SiC, Si, and SiO$_x$, which is the same order in which the wear resistances of the materials decrease.

In order to investigate a size effect in the wear behavior, we performed another series of scratch experiments using the AFM, which allowed us to decrease the contact size much further than what is achievable with the Berkovich tip. For the AFM tests we used nanocrystalline diamond tips with radii in the range of 11-27 nm (see Table 2). Figures 2a-d show scratches formed on the surfaces after 2400 cycles of scratching at a rate of 1.99 Hz over the applied load range of 120 nN to 3.5 µN (each scratch on a given surface corresponds to a different load). Magnified images of individual scratches formed at loads 2-3 µN, along with the surrounding wear debris, are shown in Figures 2e-h, and the corresponding profiles through the middle sections of the scratches are shown in Figure 2i. Surprisingly, sc-SiC and nc-SiC appear to exhibit more wear compared to Si and SiO$_x$, despite the former materials having significantly higher hardnesses.

**Table 2.** Nanocrystalline diamond AFM tip radii from each experiment.

| Sample | Set number | Radius (nm) |
|---|---|---|
| sc-SiC | I | 14.7±3.9 |
|  | II | 25.7±4.1 |
|  | III | 26.7±2.2 |
| nc-SiC | I | 14.7±3.9 |
|  | II | 25.7±4.1 |
|  | III | 26.7±2.2 |
| Si | I | 15.0±1.0 |
|  | II | 18.6±3.9 |
|  | III | 15.2±3.9 |
| $SiO_x$ | I | 14.7±3.9 |
|  | II | 14.6±2.4 |
|  | III | 11.1±1.6 |

Results shown in Figure 2j for maximum scratch depth as a function of the applied load confirm the surprising observation from the images in Figures 2a-d; namely, that at the same normal load, sc-SiC and nc-SiC show significantly more wear than what is seen for Si and $SiO_x$. The observed trend in wear resistance is present in both the elastic and the plastic regimes of deformation. Additionally, in the case of Si, its scratch depth barely exceeds the thickness of the native oxide layer, whereas both sc-SiC and nc-SiC have been scratched through the oxide layer into the bulk material even at relatively low normal loads. Although the scratches on the Si samples are shallow, as additionally shown in Figure S2, this finding is not inconsistent with results found in the literature under comparable loading and environmental conditions. For example, our wear test were conducted over a pressure range that spans both groove formation and hillock formation.[58] In the case of hillock formation, wear tracks on Si can in fact show an increase in height.[58] In addition, although Si and $SiO_2$ can have been previously shown to significant wear in humid environments[59], this high wear rate has been attributed to specific tribochemical reactions, which do not occur under dry conditions, such as those encountered in our experiments.

As shown in Figure 2j, some values of the measured scratch depth decrease with an increasing load for loads larger than 2.5 µN. This result is due to debris particles that fell into the groove and prevented the AFM tip from reaching the bottom. Since the grooves made via AFM are quite narrow, a single debris within the groove can have a significant effect on the measured depth. To verify this conclusion, we also measured the volume of the debris found on the surface as a function of the applied load. As shown in Figure S3, the wear debris volume increases with increasing load for all cases. This observation confirms that some of measurements of maximum scratch depth (Figure 2j) for the large loads underestimate the true worn volume.

To investigate any correlations between wear resistance and hardness of our samples, we plot the groove volume versus the applied load for both the tests with the nanoindenter and the tests with the AFM (Figures 3a and 3b, respectively). The groove volumes in both figures show the same qualitative trends as those of the maximum scratch depths shown in Figures 1c and 2j. By fitting equation (2) to the data shown in Figures 3a and 3b, using the measured values of hardness and sliding distance, we can calculate $k$, the wear coefficients. For the scratch experiments with the nanoindenter, $k$ values were calculated to be $1.62\times10^{-5}$, $2.79\times10^{-5}$, $5.74\times10^{-5}$, and $4.45\times10^{-5}$ for sc-SiC, nc-SiC, Si, and SiO$_x$, respectively. These values fall into the commonly reported range[60] of $10^{-2}$-$10^{-8}$. For the AFM experiments, the wear coefficients for sc-SiC, nc-SiC, Si, and SiO$_x$ were calculated to be $1.25\times10^{-2}$, $6.36\times10^{-3}$, $5.92\times10^{-6}$, and $7.72\times10^{-5}$, respectively.

In addition, to illustrate the limited applicability of the Archard's equation, in Figure 3c, we plot the relative wear resistance of each material against its hardness. We define the relative wear resistance, $R$, as:

$$R = \frac{Wx}{V} \qquad (3)$$

where $V$, $W$, and $x$ are defined as in the Archard's equation (eq 2). For the nanoindenter tests, the relative wear resistance increases with hardness, which is qualitatively consistent with the Archard's equation. This relationship is not linear (see Figure S4), which could be due to the uncertainty in the estimate of the hardness for $SiO_x$. In contrast to the nanoindenter results, the AFM data does not follow the Archard's equation even qualitatively. Specifically, the relative wear resistance is not a monotonic function of hardness. The starkest finding is that sc-SiC and nc-SiC have a lower wear resistance than Si and $SiO_x$, despite the SiC-based materials having higher hardness values. One should point out that the hardness measured in nanoindentation is representative of the hardness of the near surface layer because the nanoindentation size effect is negligible in our samples. As shown in Fig. S1b, the variation of hardness with the contact depth is much smaller than the difference in hardness values between SiC and Si.

What is the reason for the transition in the wear trends observed when the tip radius has decreased from 370 nm in the nanoindenter tests to 11 - 27 nm in the AFM tests? We hypothesize that the reason underlying the transition between wear trends reported in Figure 3 is a change in the dominant contribution to friction, even though plastic deformation is present in both sets of experiments.[52, 61] Specifically, the dominant contribution to friction in the nanoindenter tests is plowing, which arises from the materials' resistance to deformation in front of the sliding tip. In contrast, the dominant contribution to wear in the AFM tests is adhesive wear governed by interfacial shear. Because of the relatively low wear rates, the likely mechanism underlying this adhesive wear is removal of individual atoms or a small group of atoms[10-11, 62]. This type of atom-by-atom wear mechanism is consistent with the fact that in our experiments we do not find evidence of significant material transfer to the tip (see Figure S5).

To demonstrate that the plowing contribution is diminished in the AFM tests, we calculate the size of the plastic zones developed under the nanoindenter tip and under the AFM tip. The plastic zone represents the deformed volume that needs to be pushed in front of the sliding tip, which volume exerts back stress on the moving tip. The plastic zone radius, $c$, for conical and spherical indenters can be obtained from the following equation:[63-64]

$$\frac{E^*}{\sigma_y}\tan\beta = 6(1-v)\left(\frac{c}{a}\right)^3 - 4(1-2v) \qquad (4)$$

Here $E^*$ is the reduced elastic modulus, $\sigma_y$ is the yield strength, which is approximately equal to hardness/1.3, $a$ is the contact radius, and $v$ is the Poisson's ratio ($v_{sc-SiC} = 0.183$, $v_{nc-SiC} = 0.183$, $v_{Si} = 0.28$, and $v_{SiOx} = 0.28$). With the assumption of a spherical tip, we estimate $\tan\beta \approx a/R$; this value varies with the indentation depth. The calculated sizes of the plastic zones under spherical tips as a function of mean contact pressure are shown in Figure 4. First of all, one can see that the contact pressures are comparable between the nanoindenter and AFM experiments. Secondly, for the same contact pressure, the plastic zone size is much more developed under the larger nanoindenter as compared to that under the smaller AFM tip. For example, at an applied pressure of 37 GPa on sc-SiC samples, the size of the plastic zone under the nanoindenter tip ($R = 370$ nm) is ~105 nm, whereas the plastic zone under the AFM tip ($R = 14.7$ nm is) is only ~5 nm.

One should note that for the AFM scratch experiments, the max scratch depth ranged from 0 to 50 nm. Therefore some of these experiments were carried at contact depths larger than the estimated tip radius (~20 nm). In such cases, the shape of the tip might be better approximated as a cone rather than as a sphere.[65] For a conical tip, $\beta$ in equation (4) is equal to 90° minus half of the included angle of the conical indenter. As shown in Figure S6, assuming the shape of the AFM tip to be conical does not change the qualitative trends shown in Figure 4.

One should note that even though dislocation plasticity is well developed in our nanometer-sized contacts (as demonstrated for instance by Mishra *et al.*'s simulations on SiC[13]), equation (4) is still approximate. Here, we use this eqution primarily to demonstrate that the size of the plastic zone is significantly larger during nanoindenter scratching than in AFM experiments. Equation (4) relies on the calculation of the elastic stress, and plastic deformation is assumed to take place when the elastic stress reaches the value of the yield stress. Thus equation (4) does not take into account possible phase transformations in Si[66-68], but since such transformations would take place under both the nanoindenter and the AFM, our qualitative conclusion from Figure 4 still holds. Equation (4) also does not account for a nanoindentation-size effect, but as shown in Figure S1b, this effect is small in our experiments.

With the plowing contribution to friction playing a smaller role in the AFM experiments, we now need to show that the interfacial shear contribution to friction is larger for sc- and nc-SiC compared to that for Si and $SiO_x$. We made measurements of interfacial shear strength by performing wearless friction vs load AFM experiments, varying the normal load (and therefore also contact pressure) over a range that was too small to lead to permanent material deformation or wear. The measured friction force as a function of the applied load is shown in Figure 5. We imaged the scanned areas before and after the friction tests to confirm that there was no measureable wear on the surface. Two experiments on Si showed a detectable change in the friction image over the region where the friction tests were performed, which could be due to slight amounts of sample wear that were not detectable in the height images (friction images can more sensitive to surface modification than height images). Results from those tests were still included in the analysis. To obtain the interfacial shear strength from the friction vs load data, we fit the data to the Maugis-Dugdale model[44] using the Carpick-Ogletree-Salmeron (COS) equation.[45] The calculated interfacial shear strengths are shown in Figure 5b. Indeed, we find that the shear strength is higher for sc-SiC and nc-SiC than for Si and $SiO_x$ (210±48 MPa and 142±29 MPa for sc- and nc-SiC, respectively, and 72±15 MPa and 67±15 MPa for Si and $SiO_x$, respectively). This result supports the idea that in nanoscale contacts that deform plastically, a harder material can actually be less wear resistant due to the significant contribution from interfacial shear forces to friction and wear.

It is important to point out that the conditions during the wearless friction vs load tests (i.e., for the results of Figure 5) are not exactly the same as those encountered during the scratching tests. During scratching, the tip first slides along the oxide layer of the surface, but eventually the tip cuts through the oxide layer and into the bulk of the sample. However, we believe that the interfacial shear strength measured in the wearless regime is representative of the environment encountered by the tip during wear for the following reason. As opposed to the tests done on clean surfaces in ultra-high vacuum (where broken bonds may remain unpassivated), the surface bonds that become broken during our wear tests can be passivated during sliding by available oxygen atoms (e.g., oxygen from the surface oxide, oxygen from residual moisture on the surface, and/or oxygen from the air). The chemical state of the very top surface layer throughout the wear process would therefore remain similar to the unworn state.

In order to understand why sc-SiC and nc-SiC have higher shear strengths than Si and $SiO_x$, we also analyzed the roughness and the chemistry of the surfaces, as those are factors that could in principle affect the measured values of interfacial shear strength. As shown in Figure S7, we found that, over the range of roughness that our samples exhibited, the interfacial shear strengths do not depend on the root mean square (RMS) roughness. In addition, if surface roughness was a factor, the intrinsic interfacial shear strengths of sc-SiC and nc-SiC would be expected to be even higher than the values we measured due to the relatively higher roughnesses of those samples. Therefore, the differences in interfacial shear strength measured among the samples cannot be attributed to differences in roughness. Surface chemistry was analyzed using X-ray photoelectron spectroscopy (XPS), and the results are shown in Table 1. We find that the chemical nature of the oxides varies among the samples. Both sc-SiC and nc-SiC have a combination of $SiO_2$ and $Si_4C_{4-x}O_2$ on their surfaces, whereas the Si and $SiO_2$ surfaces consist solely of $SiO_2$. It is possible that the silicon oxycarbide present on sc-SiC and nc-SiC leads to higher interfacial shear strengths due to Si-C and C-C bond formation between the sample and the diamond tip during sliding. For instance, Sha *et al.*[69] found that SiC balls exhibited a stronger tendency towards adhesive wear when rubbing against a polycrystalline diamond substrate in vacuum compared to rubbing against other ceramics, and this observation was attributed to the formation of strong Si-C and C-C bonds across the sliding interface. The detrimental role of silicon oxycarbide is further supported by the observation that the shear strength of sc-SiC is higher than that of nc-SiC, and the atomic percent of the silicon oxicarbide component is twice as strong in sc-SiC as it is in nc-SiC (12% versus 6%).

One should mention that although both tips are made of diamond, the indenter is a single crystal, and the AFM tip is nanocrystalline (5 nm grain size). Within the resolution of our instruments, we did not detect any appreciable changes in either the nanoindenter tip or the AFM tip during the wear process, suggesting comparable chemical and mechanical stability (under the conditions tested) both in terms of tip degradation and accumulation of material on the tip. However, atoms at the grain boundaries and in the crystalline phase could react differently with the counter-surface material during scratching. These differences in surface termination could affect the strength of the interfacial interactions, although we are not able to detect compositional and configurational changes at that scale. Detecting atomic-level changes in composition and bonding configuration at the immediate surface of the nanoscopic tip is beyond the scope of this paper, and would require atomic-scale simulations in order to understand the effects on interfacial adhesion and shear. The switching of the trend in wear resistance of SiC and Si has not been reported before, despite the fact that other authors have used AFM tips in scratching experiment on SiC.[3, 37, 39] However, in most of the previous experiments, either the AFM tip radii[39] or the applied loads[37] were significantly higher than those in our AFM study. One exception is the study by Lantz *et al.*[3] who measured the wear behavior of SiC and Si AFM tips, where both the contact size and the applied load were in the nanoscale regime. Specifically, the tip sizes were in the range from 5 nm – 6.5 nm, and the normal load did not exceed 10 nN. The authors reported that, in this regime, amorphous SiC tips were more wear resistant than Si tips, and the wear damage was attributed to atomic attrition because the loads were too small to induce any significant plastic deformation. There are two possible reasons for the discrepancies in the conclusions from Lantz's studies and our studies. The most likely explanation lies in the low loads used by Lantz *et al*. As shown in Figure 5a, at the lower loads, we also see that sc-SiC and nc-SiC exhibit lower friction compared to Si. Another possible explanation is the different counter-surfaces used - Lantz *et al.* slid their tips against a polymeric film, whereas we slid diamond tips against our samples. Friction forces and wear rates would be expected to differ in a number of ways due to the significant differences in the mechanical and chemical properties of the differing counterfaces.

It is also worth noting that fatigue is not expected to play a major role in our experiments, as we have determined using the method proposed by Schiffmann *et al.*[70] In their experiments, breaking-off of the material due to fatigue was evidenced in discontinuous and periodic changes in the position of the AFM tip. We performed a similar analysis on our AFM data by using a triboscopic technique[71] to track the evolution of the tip's position in real time. The tracking data on sc-SiC sample is shown in Figure S8, and it can be seen that the tip's vertical position decreases continuously, which means that the removal of the material is also continuous. It should also be noted that, due to drift in the AFM system, the above method should not be used to determine the normal position of the AFM tip quantitatively. However, the qualitative trend can be used to look for signs of fatigue with reasonable confidence.

In addition to friction, we also performed a comparative study of the interfacial adhesion between the AFM tip and the four materials, the results of which are shown in Figure 6. Adhesion can be evaluated in AFM by measuring the static pull-off force between the tip and the sample. We found that the average static pull-off force measured on Si is 38.7±0.5 nN, which is higher than those of 28±16 nN for sc-SiC, 12±7 nN for nc-SiC, and 7±1 nN for $SiO_x$. The mean value of the pull-off force is higher for Si than for SiC, although the error bars are overlapping. The lower pull-off force of $SiO_2$ relative to that of SiC has also been reported in earlier studies that used Si tips.[38] We found that the dynamic pull-off forces, which can be obtained from the friction vs load data of Figure 5a, show the same qualitative trend as that of the static pull-off forces. The differences in the SiC and Si pull-off forces may come from two sources - roughness and oxide chemistry (see Table 1). It is interesting to point out that the trend in adhesion (Figure 6) are not the same as the trend in the interfacial shear strength (Figure 5b), even though both properties are largely governed by surface chemistry. This observation demonstrates that when considering the coupling of chemistry and mechanics, one needs to consider not only the magnitude of stress but also the direction of loading.

**Table 3.** Friction coefficients, $\mu$, measured in the nanoindenter and AFM experiments. $R$ is the radius of curvature.

| Sample | $\mu_{nanoindenter}$ ($R \sim 370$ nm) | $\mu_{AFM}$ ($R \sim 20$ nm) |
|---|---|---|
| sc-SiC | 0.28±0.04 | 0.34±0.08 |
| nc-SiC | 0.32±0.01 | 0.37±0.05 |
| Si | 0.23±0.01 | 0.22±0.03 |
| SiO$_x$ | 0.27±0.01 | 0.13±0.04 |

In addition to wear resistance, we also analyzed the friction forces during sliding and calculated friction coefficients, $\mu$. $\mu$ is defined as the slope of the friction force vs normal load curve (provided that this relationship is linear). The friction force vs load data are reported in Figures 7a and 7b for the nanoindenter and AFM experiments, respectively, and the friction coefficients are summarized in Table 3. We find that, in both experiments, the average friction coefficients are higher for sc-SiC and nc-SiC than they are for Si and SiO$_x$, although with the data from the nanoindenter, the error bars for sc-SiC and Si are overlapping. Because harder materials typically have a lower coefficient of friction when in the regime of plastic deformation, these results demonstrate that when the contact size is reduced, the interfacial shear strength can play a dominant role, even when the degree of plastic deformation is significant. However, the switching trend that was observed in the wear resistance is not seen when comparing the friction coefficients measured by the nanoindenter and the AFM. Future modeling work will be aimed at quantifying the relative contributions of plowing and interfacial shear to wear for these hard materials as the size of the contact is reduced to the low end of the nanoscale.

**CONCLUSION**

We investigated the single-asperity wear behavior of sc-SiC and nc-SiC via scratching experiments with both a nanoindenter and an AFM in dry environments, and we compared the results to those obtained on Si surfaces with oxide layers. We found that, over the same range of contact pressures, the wear resistance of SiC can be switched relative to that of Si by changing the size of the contact, thereby showing that wear resistance can become uncorrelated with hardness in nanoscopic mechanical contacts even when the plastic zone is well developed. Thus SiC, a mechanically hard material that is often considered for use in protective coatings, can be less resistant to wear than the significantly softer Si due to a size effect at the nanoscale. We attribute the switching behavior to a transition from a wear regime dominated plowing, with larger tips and higher loads, to a regime dominated by interfacial shear, with smaller tips and lower loads. Our studies demonstrate that when considering wear resistance of materials, it is important to take into account not only the hardness and the surface chemistry, but also the applied load and the size of the contact.

**Acknowledgment**

C. Tangpatjaroen, D. Grierson and I. Szlufarska gratefully acknowledge support from the Department of Energy, Basic Energy Sciences grant #DE-FG02-08ER46493. S. Shannon acknowledges support from the Department of Energy, Nuclear Energy University Partnership grant #AC07-05ID14517. In addition, we also would like to thank J. Jacobs, and Dr. J. Last for their help in setting up experimental procedure.

**Supporting Information Available:**

Eight supplementary figures are presented. These materials are available free of charge via the Internet at http://pubs.acs.org.

**References**

1. Heinz, D. B.; Hong, V. A.; Ahn, C. H.; Ng, E. J.; Yang, Y.; Kenny, T. W., Experimental Investigation Into Stiction Forces and Dynamic Mechanical Anti-Stiction Solutions in Ultra-Clean Encapsulated MEMS Devices. *J. Microelectromech. Syst.* **2016,** *25*, 469-478.
2. Mathew, C. M., *Tribology on the Small Scale : A Bottom Up Approach to Friction, Lubrication, and Wear*. Oxford University Press: New York, 2008.
3. Lantz, M. A.; Gotsmann, B.; Jaroenapibal, P.; Jacobs, T. D. B.; O'Connor, S. D.; Sridharan, K.; Carpick, R. W., Wear-Resistant Nanoscale Silicon Carbide Tips for Scanning Probe Applications. *Adv. Funct. Mater.* **2012,** *22*, 1639-1645.
4. Tambe, N. S.; Bhushan, B., Scale Dependence of Micro/Nano-Friction and Adhesion of MEMS/NEMS Materials, Coatings and Lubricants. *Nanotechnology* **2004,** *15*, 1561.
5. Kumar, D. D.; Kumar, N.; Kalaiselvam, S.; Dash, S.; Jayavel, R., Micro-Tribo-Mechanical Properties of Nanocrystalline TiN Thin Films for Small Scale Device Applications. *Tribol. Int.* **2015,** *88*, 25-30.
6. Mishra, M.; Tangpatjaroen, C.; Szlufarska, I., Plasticity-Controlled Friction and Wear in Nanocrystalline SiC. *J. Am. Ceram. Soc.* **2014,** *97*, 1194-1201.
7. Luan, B.; Robbins, M. O., The Breakdown of Continuum Models for Mechanical Contacts. *Nature* **2005,** *435*, 929-932.
8. Mo, Y.; Turner, K. T.; Szlufarska, I., Friction Laws at the Nanoscale. *Nature* **2009,** *457*, 1116-1119.

9. Deng, Z.; Smolyanitsky, A.; Li, Q.; Feng, X.-Q.; Cannara, R., Adhesion-Dependent Negative Friction Coefficient on Chemically Modified Graphite at the Nanoscale. *Nat. Mater.* **2012,** *11*, 1032-1037.
10. Jacobs, T. D. B.; Carpick, R. W., Nanoscale Wear as a Stress-Assisted Chemical Reaction. *Nat. Nanotechnol.* **2013,** *8*, 108-112.
11. Gotsmann, B.; Lantz, M. A., Atomistic Wear in a Single Asperity Sliding Contact. *Phys. Rev. Lett.* **2008,** *101*, 125501.
12. Patten, J.; Gao, W.; Yasuto, K., Ductile Regime Nanomachining of Single-Crystal Silicon Carbide. *J. Manuf. Sci. Eng.* **2004,** *127*, 522-532.
13. Mishra, M.; Szlufarska, I., Dislocation Controlled Wear in Single Crystal Silicon Carbide. *J. Mater. Sci.* **2013,** *48*, 1593-1603.
14. Schirmeisen, A., Wear: One Atom after the Other. *Nat Nano* **2013,** *8*, 81-82.
15. Li, X.; Bhushan, B., Micro/nanomechanical Characterization of Ceramic Films for Microdevices. *Thin Solid Films* **1999,** *340*, 210-217.
16. Gradinaru, G.; Sudarshan, T. S.; Gradinaru, S. A.; Mitchell, W.; Hobgood, H. M., Electrical Properties of High Resistivity 6H–SiC under High Temperature/High Field Stress. *Appl. Phys. Lett.* **1997,** *70*, 735-737.
17. Brütsch, R., Chemical Vapour Deposition of Silicon Carbide and Its Applications. *Thin Solid Films* **1985,** *126*, 313-318.
18. Wright, N. G.; Horsfall, A. B.; Vassilevski, K., Prospects for SiC Electronics and Sensors. *Mater. Today* **2008,** *11*, 16-21.
19. Knotek, O.; Löffler, F.; Wolkers, L., Diamond 1992 Proceedings of the Third International Conference on the New Diamond Science and Technology Amorphous SiC PVD coatings. *Diamond Relat. Mater.* **1993,** *2*, 528-530.
20. Stoldt, C. R.; Fritz, M. C.; Carraro, C.; Maboudian, R., Micromechanical Properties of Silicon-Carbide Thin Films Deposited using Single-Source Chemical-Vapor Deposition. *Appl. Phys. Lett.* **2001,** *79*, 347-349.
21. Wijesundara, M. B. J.; Stoldt, C. R.; Carraro, C.; Howe, R. T.; Maboudian, R., Nitrogen Doping of Polycrystalline 3C–SiC Films Grown by Single-Source Chemical Vapor Deposition. *Thin Solid Films* **2002,** *419*, 69-75.
22. Jiang, L.; Cheung, R., A Review of Silicon Carbide Development in MEMS Applications. *Comput. Mater. Sci.* **2009,** *2*, 227-242.
23. Roy, S.; DeAnna, R. G.; Zorman, C. A.; Mehregany, M., Fabrication and Characterization of Polycrystalline SiC Resonators. *IEEE Trans. Electron Devices* **2002,** *49*, 2323-2332.
24. Zorman, C. A.; Parro, R. J., Micro- and Nanomechanical Structures for Silicon Carbide MEMS and NEMS. *Phys. Status Solidi B* **2008,** *245*, 1404-1424.
25. Yang, Y. T.; Ekinci, K. L.; Huang, X. M. H.; Schiavone, L. M.; Roukes, M. L.; Zorman, C. A.; Mehregany, M., Monocrystalline Silicon Carbide Nanoelectromechanical Systems. *Appl. Phys. Lett.* **2001,** *78*, 162-164.
26. Gao, D.; Carraro, C.; Howe, R. T.; Maboudian, R., Polycrystalline Silicon Carbide as a Substrate Material for Reducing Adhesion in MEMS. *Tribol. Lett.* **2006,** *21*, 226-232.
27. Ashurst, W. R.; Wijesundara, M. B. J.; Carraro, C.; Maboudian, R., Tribological Impact of SiC Encapsulation of Released Polycrystalline Silicon Microstructures. *Tribol. Lett.* **2004,** *17*, 195-198.

28. Wang, Y.; Jiang, S.; Wang, M.; Wang, S.; Xiao, T. D.; Strutt, P. R., Abrasive Wear Characteristics of Plasma Sprayed Nanostructured Alumina/Titania Coatings. *Wear* **2000,** *237*, 176-185.
29. Deng, X. Y.; Wang, X. H.; Gui, Z. L.; Li, L. T.; Chen, I. W., Grain-Size Effects on the Hardness of Nanograin $BaTiO_3$ Ceramics. *J. Electroceram.* **2008,** *21*, 238-241.
30. Szlufarska, I.; Nakano, A.; Vashishta, P., A Crossover in the Mechanical Response of Nanocrystalline Ceramics. *Science* **2005,** *309*, 911-914.
31. Mo, Y.; Szlufarska, I., Simultaneous Enhancement of Toughness, Ductility, and Strength of Nanocrystalline Ceramics at High Strain-Rates. *Appl. Phys. Lett.* **2007,** *90*, 181926.
32. Liao, F.; Girshick, S. L.; Mook, W. M.; Gerberich, W. W.; Zachariah, M. R., Superhard Nanocrystalline Silicon Carbide Films. *Appl. Phys. Lett.* **2005,** *86*, 171913.
33. Miyoshi, K.; Buckley, D. H., XPS, AES and Friction Studies of Single-Crystal Silicon Carbide. *Appl. Surf. Sci.* **1982,** *10*, 357-376.
34. Yamamoto, Y.; Ura, A., Influence of Interposed Wear Particles on the Wear and Friction of Silicon Carbide in Different Dry Atmospheres. *Wear* **1992,** *154*, 141-150.
35. Takadoum, J.; Zsiga, Z.; Roques-Carmes, C., Wear Mechanism of Silicon Carbide: New Observations. *Wear* **1994,** *174*, 239-242.
36. Li, X.; Wang, X.; Bondokov, R.; Morris, J.; An, Y. H.; Sudarshan, T. S., Micro/Nanoscale Mechanical and Tribological Characterization of SiC for Orthopedic Applications. *J. Biomed. Mater. Res., Part B* **2005,** *72B*, 353-361.
37. Pöhlmann, K.; Bhushan, B.; Zum Gahr, K.-H., Effect of Thermal Oxidation on Indentation and Scratching of Single-Crystal Silicon Carbide on Microscale. *Wear* **2000,** *237*, 116-128.
38. Zum Gahr, K. H.; Blattner, R.; Hwang, D. H.; Pöhlmann, K., Micro- and Macro-Tribological Properties of SiC Ceramics in Sliding Contact. *Wear* **2001,** *250*, 299-310.
39. Sundararajan, S.; Bhushan, B., Micro/nanotribological Studies of Polysilicon and SiC Films for MEMS Applications. *Wear* **1998,** *217*, 251-261.
40. Jamison, L.; Zheng, M.-J.; Shannon, S.; Allen, T.; Morgan, D.; Szlufarska, I., Experimental and Ab Initio Study of Enhanced Resistance to Amorphization of Nanocrystalline Silicon Carbide under Electron Irradiation. *J. Nucl. Mater.* **2014,** *445*, 181-189.
41. Cannara, R. J.; Eglin, M.; Carpick, R. W., Lateral Force Calibration in Atomic Force Microscopy: A New Lateral Force Calibration Method and General Guidelines for Optimization. *Rev. Sci. Instrum.* **2006,** *77*, 053701.
42. Ogletree, D. F.; Carpick, R. W.; Salmeron, M., Calibration of Frictional Forces in Atomic Force Microscopy. *Rev. Sci. Instrum.* **1996,** *67*, 3298-3306.
43. Cannara, R. J.; Brukman, M. J.; Carpick, R. W., Cantilever tilt compensation for variable-load atomic force microscopy. *Rev. Sci. Instrum.* **2005,** *76*, 053706.
44. Maugis, D., Adhesion of Spheres: The JKR-DMT Transition Using a Dugdale Model. *J. Colloid Interface Sci.* **1992,** *150*, 243-269.
45. Carpick, R. W.; Ogletree, D. F.; Salmeron, M., A General Equation for Fitting Contact Area and Friction vs Load Measurements. *J. Colloid Interface Sci.* **1999,** *211*, 395-400.
46. Stone, D. S.; Yoder, K. B.; Sproul, W. D., Hardness and Elastic Modulus of TiN Based on Continuous Indentation Technique and New Correlation. *J. Vac. Sci. Technol., A* **1991,** *9*, 2543-2547.

47. Oliver, W. C.; Pharr, G. M., An Improved Technique for Determining Hardness and Elastic Modulus using Load and Displacement Sensing Indentation Experiments. *J. Mater. Res.* **1992,** *7*, 1564-1583.
48. Stone, D. S., Elastic Rebound between an Indenter and a Layered Specimen I: Model. *J. Mater. Res.* **1998,** *13*, 6.
49. Tapily, K.; Jakes, J. E.; Gu, D.; Baumgart, H.; Elmustafa, A. A., Nanomechanical Study of Amorphous and Polycrystalline ALD $HfO_2$ Thin Films. *Int. J. Surf. Sci. Eng.* **2011,** *5*, 193-204.
50. Yoder, K. B.; Stone, D. S.; Hoffman, R. A.; Lin, J. C., Elastic Rebound Between an Indenter and a Layered Specimen. II. Using Contact Stiffness to Help Ensure Reliability of Nanoindentation Measurements. *J. Mater. Res.* **1998,** *13*, 3214-3220.
51. Oliver, W. C.; Pharr, G. M., Measurement of Hardness and Elastic Modulus by Instrumented Indentation: Advances in Understanding and Refinements to Methodology. *J. Mater. Res.* **2004,** *19*, 3-20.
52. Bhushan, B., *Introduction to Tribology*. 2 ed.; John Wiley & Sons: New York, 2013.
53. Meyers, M. A.; Mishra, A.; Benson, D. J., Mechanical Properties of Nanocrystalline Materials. *Prog. Mater Sci.* **2006,** *51*, 427-556.
54. Suryanarayana, C.; Mukhopadhyay, D.; Patankar, S. N.; Froes, F. H., Grain Size Effects in Nanocrystalline Materials. *J. Mater. Res.* **1992,** *7*, 2114-2118.
55. Kruschov, M. M., Resistance of Metals to Wear by Abrasion, as Related to Hardness. In *Proceedings of Conference on Lubrication and Wear: Instn Mech. Engrs*, London, UK, 1957, pp 655-659.
56. Zum Gahr, K. H., *Microstructure and Wear of Materials*. 1 ed.; Elsevier: Amsterdam, Halland, 1987; p 559.
57. Archard, J. F., Contact and Rubbing of Flat Surfaces. *J. Appl. Phys.* **1953,** *24*, 981-988.
58. Yu, J. X.; Qian, L. M.; Yu, B. J.; Zhou, Z. R., Nanofretting behaviors of monocrystalline silicon (1 0 0) against diamond tips in atmosphere and vacuum. *Wear* **2009,** *267*, 322-329.
59. Yu, J.; Kim, S. H.; Yu, B.; Qian, L.; Zhou, Z., Role of Tribochemistry in Nanowear of Single-Crystalline Silicon. *ACS Appl. Mater. Interfaces* **2012,** *4*, 1585-1593.
60. Rabinowicz, E., The Wear Coefficient—Magnitude, Scatter, Uses. *J. Lubr. Technol.* **1981,** *103*, 188-193.
61. Lafaye, S.; Gauthier, C.; Schirrer, R., The Ploughing Friction: Analytical Model with Elastic Recovery for a Conical Tip with a Blunted Spherical Extremity. *Tribol. Lett.* **2006,** *21*, 95-99.
62. Bhaskaran, H.; Gotsmann, B.; Sebastian, A.; Drechsler, U.; Lantz, M. A.; Despont, M.; Jaroenapibal, P.; Carpick, R. W.; Chen, Y.; Sridharan, K., Ultralow nanoscale wear through atom-by-atom attrition in silicon-containing diamond-like carbon. *Nat Nano* **2010,** *5*, 181-185.
63. Johnson, K. L., *Contact Mechanics*. Cambridge University Press: 1985.
64. Woodcock, C. L.; Bahr, D. F., Plastic Zone Evolution Around Small Scale Indentations. *Scripta Mater.* **2000,** *43*, 783-788.
65. Bonilla, M. R.; Stokes, J. R.; Gidley, M. J.; Yakubov, G. E., Interpreting Atomic Force Microscopy Nanoindentation of Hierarchical Biological Materials using Multi-Regime Analysis. *Soft Matter* **2015,** *11*, 1281-1292.
66. Kiran, M. S. R. N.; Haberl, B.; Bradby, J. E.; Williams, J. S., Chapter Five - Nanoindentation of Silicon and Germanium. In *Semiconductors and Semimetals*, Lucia Romano, V. P.; Chennupati, J., Eds. Elsevier: 2015; Vol. Volume 91, pp 165-203.

67. Goel, S.; Faisal, N. H.; Luo, X.; Yan, J.; Agrawal, A., Nanoindentation of polysilicon and single crystal silicon: molecular dynamics simulation and experimental validation. *J. Phys. D: Appl. Phys.* **2014,** *47*, 275304.
68. Zhang, Z.; Guo, D.; Wang, B.; Kang, R.; Zhang, B., A novel approach of high speed scratching on silicon wafers at nanoscale depths of cut. *Scientific Reports* **2015,** *5*, 16395.
69. Sha, X.; Yue, W.; Zhao, Y.; Lin, F.; Wang, C., Effect of Sliding Mating Materials on Vacuum Tribological Behaviors of Sintered Polycrystalline Diamond. *Int. J. Refract. Met. Hard Mater.* **2016,** *54*, 116-126.
70. Schiffmann, K., Microwear Experiments on Metal-Containing Amorphous Hydrocarbon Hard Coatings by AFM: Wear Mechanisms and Models for the Load and Time Dependence. *Wear* **1998,** *216*, 27-34.
71. Loubet, J. L.; Belin, M.; Durand, R.; Pascal, H., Triboscopic Description of Local Wear Phenomena under an AFM Tip. *Thin Solid Films* **1994,** *253*, 194-198.

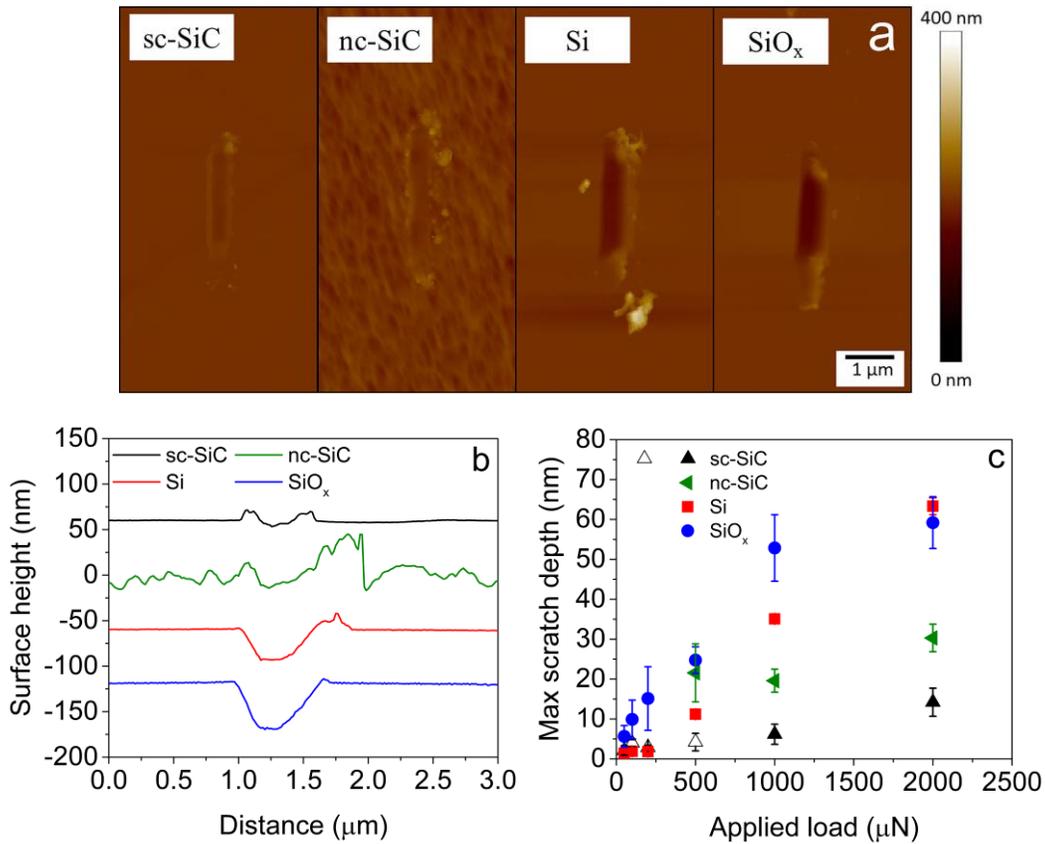

**Figure 1.** (Color online) Results from scratch testing of sc-SiC, nc-SiC, Si, and SiO$_x$ for 500 cycles by a Berkovich diamond tip with a radius of 370 nm. a) AFM image of the scratches formed under a load of 1 mN. The scale bar represents the height of the surface features. b) Cross-sectional profiles of the scratches taken through the middle of each scratch under a normal load of 1 mN. c) Maximum scratch depth versus the load applied. Open and solid symbols represent points for which the pressure in the contact is lower than and equal to the materials' hardnesses, respectively.

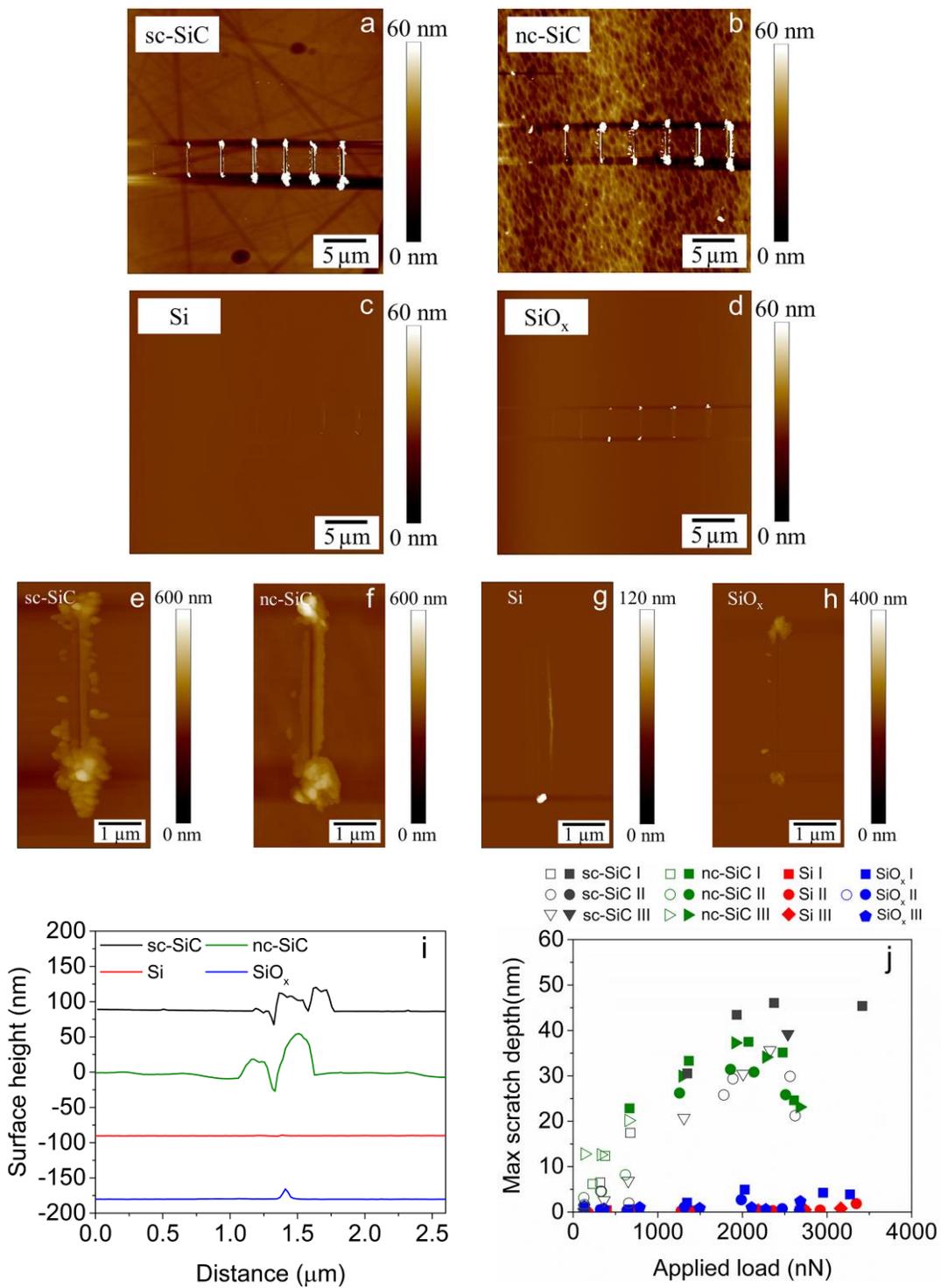

**Figure 2.** (Color online) AFM images of scratch tests performed over a range of applied normal loads using a nanocrystalline diamond AFM tip sliding on a) sc-SiC, b) nc-SiC, c) Si, and d) SiO$_x$ after 2400 reciprocal cycles. Each scratch is 3 µm long and corresponds to a different normal load. Wear debris on e) sc-SiC, f) nc-SiC, g) Si, and h) SiO$_x$. i) Cross-sectional surface profile of the scratches under the normal load of 2-3 mN. j) Maximum scratch depth versus the applied load. Open and solid symbols represent measurements with the applied pressure lower than and equal to the material's hardness, respectively. The labels I, II, and III refer to tests performed with tips of different radii, as specified in Table 2.

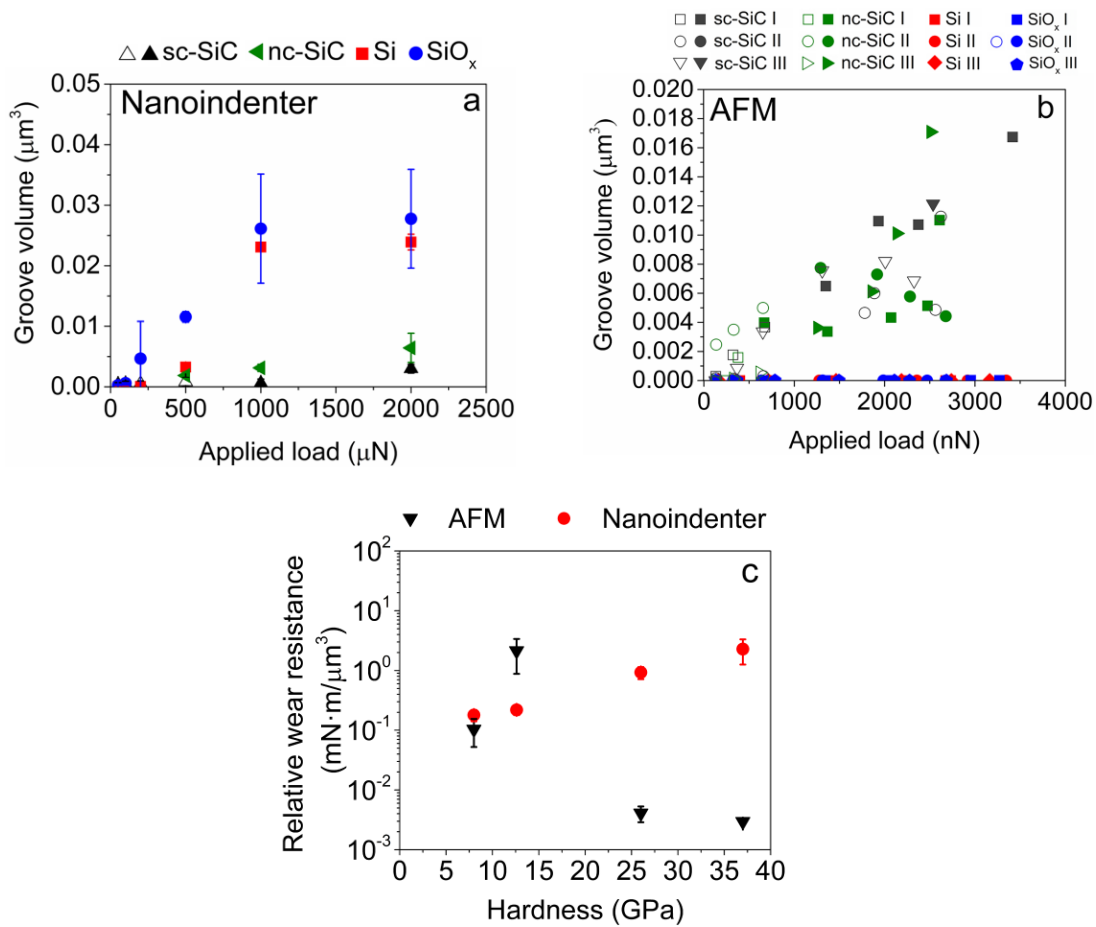

**Figure 3.** (Color online) Groove volume as a function of normal load for a) nanoindentater and b) AFM scratch experiments. Open and solid symbols correspond to mean contact pressures that are lower than and equal to hardness, respectively. c) Relative wear resistance of the four materials versus hardness. The same data is plotted on a linear scale in Figure S4.

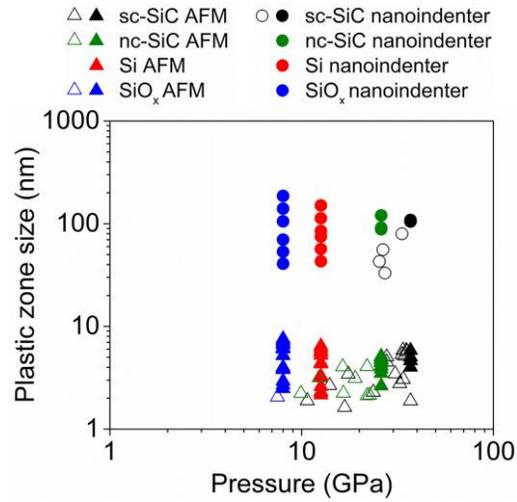

**Figure 4.** (Color online) The size of the plastic zone as a function of the mean contact pressure for the nanoindenter (circles) and AFM (triangles) scratch tests. Open and solid symbols represent measurements for which the mean contact pressure was lower than or equal to the hardness of a given material, respectively.

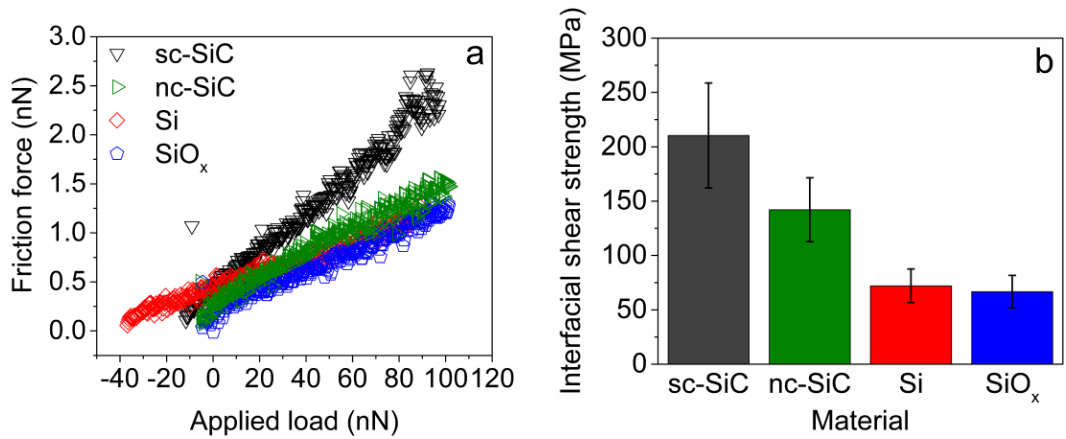

**Figure 5.** (Color online) Wearless friction tests. a) Friction force vs applied load. The results shown correspond to RMS roughness of 0.307 nm, 0.343 nm, 0.0682 nm, and 0.0762 nm, for sc-SiC, nc-SiC, Si, and SiO$_x$, respectively. b) Interfacial shear strengths obtained from fitting the data in a) to the COS equation.

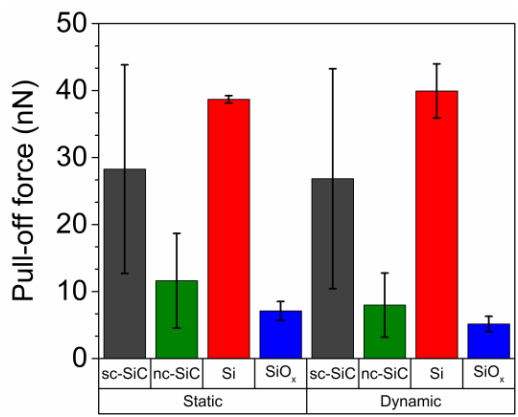

**Figure 6.** (Color online) Static and dynamic pull-off forces measured via AFM.

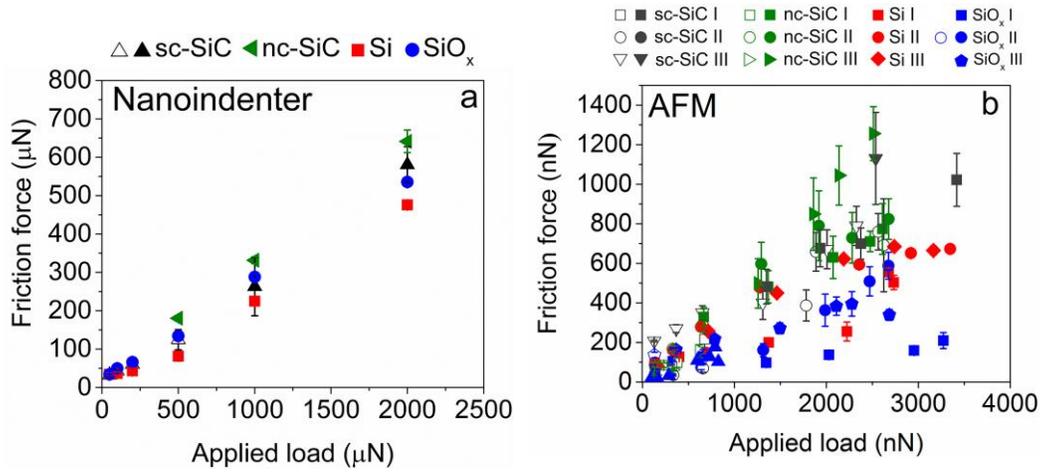

**Figure 7.** (Color online) Friction force versus the applied load from a) nanoindentater scratch tests with a Berkovich diamond tip and b) AFM scratch tests with a nanocrystalline diamond tip. Open and solid symbols represent measurements during which the contact pressure is lower than or equal to hardness of a given material, respectively.

**Table of Contents (TOC)**

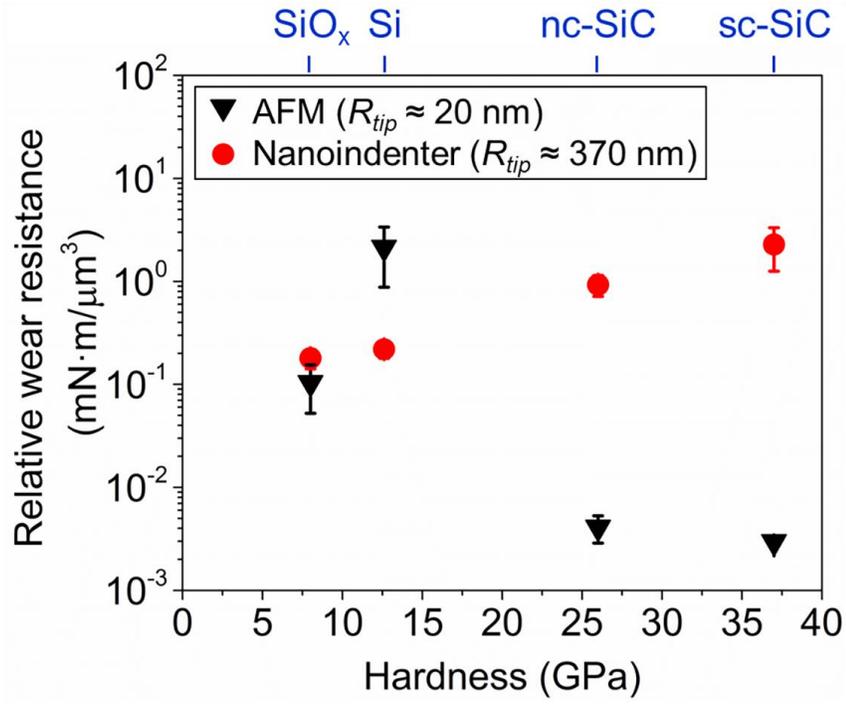